\documentclass[fleqn,usenatbib]{mnras}

\usepackage{newtxtext,newtxmath}

\usepackage[T1]{fontenc}
\usepackage{placeins}
\DeclareRobustCommand{\VAN}[3]{#2}
\let\VANthebibliography\thebibliography
\def\thebibliography{\DeclareRobustCommand{\VAN}[3]{##3}\VANthebibliography}

\UseRawInputEncoding
\usepackage{graphicx}
\usepackage{subcaption}
\usepackage{amsmath}	

\title[Short title, max. 45 characters]{Multi-wavelength temporal and spectral analysis of Blazar S5 1803+78 }

\author[S. Priya et al.]{
Shruti Priya,$^{1}$\thanks{E-mail: shruti62442@iitkgp.ac.in}
Raj Prince,$^{2}$
Aditi Agarwal,$^{3}$
Debanjan Bose,$^{4}$\thanks{E-mail: debanjan.bose@bose.res.in}
Aykut \"Ozd\"onmez,$^{5}$
and Erg\"un Ege$^{6}$
\\
$^{1}$Department of Physics, Indian Institute of Technology Kharagpur, 721302, India\\
$^{2}$Center for Theoretical Physics, Polish Academy of Sciences, Lotnik\'{o}w 32/46, Warsaw, Poland\\
$^{3}$Raman Research Institute, C. V. Raman Avenue, Sadashivanagar, Bengaluru - 560 080, India\\
$^{4}$Department of Astrophysics and Cosmology, S N Bose National Centre for Basic Sciences, Kolkata, 700106, India \\
$^{5}$Ataturk University, Faculty of Science,  Department of Astronomy and Space Science, 25240, Yakutiye, Erzurum \\
$^{6}$Istanbul University, Faculty of Science, Department of Astronomy and Space Sciences, 34116, Beyazıt, Istanbul, Turkey
}

\date{Accepted XXX. Received YYY; in original form ZZZ}

\pubyear{2022}

\begin{document}
\label{firstpage}
\pagerange{\pageref{firstpage}--\pageref{lastpage}}
\maketitle

\begin{abstract}
Blazars are a class of AGN, one of their jets is pointed towards the earth. Here, we report about the multi-wavelength study for blazar S5 1803+78 between MJD 58727 to MJD 59419. We analysed $\gamma$-ray data collected by Fermi-LAT,  X-ray data collected by Swift-XRT \& NuSTAR, optical photons detected by Swift-UVOT \& TUBITAK observatory in Turkey. Three flaring states are identified by analysing the $\gamma$-ray light curve. A day scale variability is observed throughout the flares with the similar rise and decay times suggesting a compact emission region located close to the central engine. 
Cross-correlation studies are carried out between $\gamma$-ray, radio, and X-ray bands, and no significant correlation is detected. The $\gamma$-ray and optical emission are significantly correlated with zero time lag suggesting a co-spatial origin of them. A significant positive correlation between the R-I index and the V magnitude is observed.
The broadband spectral energy distributions (SEDs) modeling was performed for all the flaring episodes as well as for one quiescent state for comparison.
SEDs are best fitted with the synchrotron-self Compton (SSC) model under a one-zone leptonic scenario. The SED modeling shows that to explain the high flaring state strong Doppler boosting is required.
\end{abstract}




\section{Introduction}

Active galactic nuclei (AGN) are the center of an active galaxy, generally believed to have a supermassive black hole (SMBH) at the center surrounded by an accretion disk (\citealt{krolik_book1999}) and a bi-polar highly relativistic jet perpendicular to the plane of the disk. Various studies done on AGN suggest that the central part of the AGN is covered by the dense molecular gas cloud known as a molecular torus or dusty torus. How the relativistic jet is formed and the acceleration of charged particles to very high energy is still an open question to the community. The recent EHT observations of M87 have revealed the existence of SMBH at the center of an active galaxy (\citealt{EHT2019_shadow, EHT2019}) which is proposed to power the jet. AGNs are classified into various types based on how they are viewed (\citealt{Antonucci1993, Urry1995}). AGN seen along the jet axis within a few degrees are classified as blazars (\citealt{Urry1995}). Due to the small viewing angle and high Doppler boosting the broadband emission produced in the blazar is boosted along the jet axis and emission is observed to be highly variable and polarized. The main temporal characteristics of blazars are those which show fast-flux variability order of minutes to hours (\citealt{Heidt1996, Ulrich1997}) and spectacular flares on the time scale of the week to months across the wavebands ranging from low frequency radio to very high energy $\gamma$-ray. The broadband spectral energy distribution (SED) of blazar show two hump structure in the low and high energy part of the SED, respectively. Studies done in the past confirmed that the low energy hump can be produced by the synchrotron emission by charged leptons moving in the entangled jet magnetic field. On the other hand, the origin of the high energy hump is debatable. To explain the high energy hump, two physical models have been put forward namely, leptonic and hadronic involving leptons and hadrons as charged particles as names suggest. The main physical mechanism under consideration in the leptonic model is the inverse-Compton scattering of low energy photons by the relativistic electrons/positrons. The source of low energy photons could be internal to jet (i.e. synchrotron photon) and the process is known as synchrotron self-Compton (SSC; \citealt{Konigl1981, Marsche1985, Ghisellini1989}), or external to jet from the broad-line region and/or dusty torus, the process is identified as external-Compton (EC) (\citealt{Begelman1987}). In the hadronic scenario, the possible physical processes are proton-synchrotron, proton-proton, and proton-photon interactions (\citealt{Dermer1993, Boettcher1997}) responsible for the high energy peak of the SED. Blazars are further classified as flat-spectrum radio quasars (FSRQ) and BL Lacerate (BL Lac) objects based on various observational properties. Initially, it is found that the FSRQ show broad and strong optical emission lines in their spectra whereas weak or no emission lines are detected in BL Lac. Later, the classification is done based on the location of the synchrotron peak in broadband SED which included FSRQs and the different classes within the BL Lacs. For FSRQs the synchrotron peak is located at frequencies ($\nu$) $<$ 10$^{14}$ Hz, for LBLs i.e. low frequency BL Lacs this peak is located between 10$^{14}$ $<$ $\nu$ $<$ 10$^{15}$ Hz (\citealt{Fan2016}), synchrotron peak for  intermediate BL Lacs (IBL) located between 10$^{15}$ $<$ $\nu$ $<$ 10$^{16}$ Hz, for high BL Lacs (HBL) between 10$^{16}$ $<$ $\nu$ $<$ 10$^{18}$ Hz and for extreme HBL (EHBL) above $\nu$ $>$ 10$^{18}$ Hz (\citealt{Abdo_2010}).

Blazars are the extreme accelerator in the universe and an ideal place to test extreme physics. The Fermi-acceleration (diffusive shock acceleration) mechanism has been proposed initially as an efficient process to accelerate the charged particle to very high energy and which is the source of fast flux variability often seen in blazar. However, in a few sources minutes scale variability has been observed which is very challenging to explain with shock acceleration, and hence magnetic re-connection model has been put forward (\citealt{Shukla2020}).

\citet{Nesci} have monitored this source in optical for a long period of time between 1996 to 2011. They searched for the periodicity in the optical light curves but did not see any clear evidence but they suggest 1300 days periodicity between the bright outburst. Their study also commented on the no clear trend of color index variation with the flux on a longer time scale however they claim the bluer trend in the rising phase and redder trend in the decaying phase of one of the optical flares. During their optical monitoring, they also observed two-$\gamma$-ray flare by Fermi-GST (now known as Fermi-LAT) and one of them coincides with the optical brightening. Having the simultaneous observation in optical and gamma-ray during the flaring episode, they modeled the flare with a one-zone SSC model which describes the source behavior.

Blazar S5 1803+78 is located at redshift, z=0.684 (\citealt{Lawrence_1996}) and classified as BL Lac type object (\citealt{Biermann1981}). Under the AGN unification scheme by \citet{Urry1995} it is identified as low-synchrotron peak (LSP) blazar. Long optical and radio variations are studied in \citet{2002AJ....124...53N, Nesci}.
Based on 30 years long VLBI observations, \citet{kun2018} have detected a periodicity of 6-years which is linked to the helical jet motion.  In a recent study by \citet{Nesci2021} they studied the morphological change in the radio structure after the bright $\gamma$-ray flares and show that two new components were originated from the core and moved outwards at the time of bright $\gamma$-ray flares.

In this paper, we present the broadband study of the source collecting the data from radio to $\gamma$-ray. The broadband SED modeling is performed to understand the multi-wavelength flaring events in the source. In section 2, we describe multi-waveband observations and data analysis methods, multi-waveband light curves are discussed in section 3. In section 4 we have explained gamma-ray spectral data, followed by multi-wavelength modeling in section 5. We discuss our results in section 6 and finally summary in section 7.

\section{Multi-waveband observations and data analysis}

The following section describes how the \textit{Fermi}-LAT, \textit{Swift}-XRT, \textit{Swift}-UVOT, NuSTAR, and Optical data was analyzed to obtain the multi-waveband light curve. Radio data from OVRO was used for the correlation study. Section \ref{sec3} describes the analysis carried out on the multi-frequency data to identify the flaring regions. The multi-waveband SED was also obtained for the data and that has been described in Section \ref{sec4}.

\subsection{High energy $\gamma$-ray observations of \textit{Fermi}-LAT}
The $\gamma$-ray data used were collected by the Fermi Large Area Telescope (LAT) instrument onboard the Fermi $\gamma$-ray Space Telescope in the 100 MeV to 500 GeV energy range. The Fermi-LAT is a pair conversion $\gamma$-ray detector in the near-Earth orbit since June 2008. It has a large FoV of about 2.4 sr and single photon resolution of < $3.5^{\circ}$ at 100 MeV energy and improves to less than $0.6^{\circ}$ for energies greater than 1 GeV \citep{fermi2009}. The analysis is done with the help of standard analysis procedure provided by the ScienceTools\footnote{https://fermi.gsfc.nasa.gov/ssc/data/analysis/scitools/python\_tutorial.html}. The time range for the Fermi-LAT data is from MJD 58727 to MJD 59419. Events were extracted from a circular region of interest (ROI) with a   $15^{\circ}$ radius centered around the source position (RA: 270.19, DEC: 78.468). The photon data files were filtered using "$evclass=128$" and "$evtype=3$" as recommended in the fermitools documentation. Good time intervals were selected using "$(DATA\_{QUAL}>0) \&\& (LAT\_{CONFIG}==1)$". A maximum zenith angle cut $>$ 90 is applied to reduce the contamination from Earth limb $\gamma$-rays. The source model xml file is generated using the 4FGL sources catalog \citep{4fgl2020} and the unbinned likelihood is performed using "gtlike" to optimize the source parameters \citep{cash1979,mattox1996}. The isotropic and diffuse background models, "$iso\_P8R3\_SOURCE\_V3\_v1.txt$" and "$gll\_iem\_v07.fits$", were used in the process. "gtlike" also returns the significance of all the sources within the ROI in terms of test statistics (TS).
The $\gamma$-ray light curve and $\gamma$-ray SED data points for S5 1803+78 were obtained.
Multiple light curves were computed using different time bins using the 4FGL model with 221 point sources. Three of the sources are within $5^{\circ}$ radius of S51803+78. The unbinned likelihood was performed using gtlike with 11 free parameters. 
The log parabola spectral model was used for our source of interest as described below
\begin{gather}
\frac{dN(E)}{dE} = N_0 \times \Big(\frac{E}{E_0}\Big)^{-(\alpha +\beta log(E/E_0))}
\end{gather}
Here $N_0$ is the pre-factor, $E_0$ is the scaling factor, $\alpha$ is the spectral index and $\beta$ is the curvature index. The light curve in Figure~\ref{fig:LC} was produced using a bin size of 7 days. The test statistic was $> 9$ for all the time bins and reached values of around 2000 for the flaring period.

\subsection{X-ray observations}
{\bf \textit{Swift}-XRT:}

X-ray data from \textit{Swift}-XRT was analyzed for the energy range of 0.3 to 8 keV. A total of 55 observations are available between MJD 58696 to MJD 58957. These data are simultaneous with Fermi-LAT data in gamma-ray but cover the first half of the gamma-ray selected period. The X-ray data are exactly covering the pre-flaring region defined from the gamma-ray light curve.
X-ray data for the rest of the gamma-ray light curve is not available.

HEASoft 6.29 and XSPEC 12 were used to analyze the X-ray data. 
The package \texttt{xrtpipeline} was used to get the clean events from the raw data. The source and background events were obtained from circular regions of 20 and 40 pixels(1 pixel ~ 2.36") in radii around the source and away from the source. 
The spectrum files were obtained using a tool \texttt{xselect}. The redistribution matrix-response files (RMFs) were obtained from HEASARC calibration database. Ancillary response files were produced using \texttt{xrtmkarf}. The tool \texttt{grppha} was used to tie the source, background, response, and ancillary response files together and binned to get a minimum of 20 counts. The grouped and binned spectrum file is eventually used in \texttt{XSPEC} for modeling and extracting the flux and index. The fit was done with a basic power law, $F(E)=KE^{-\Gamma_x}$. The neutral column density for this source was fixed to $N_H = 3.64 \times 10^{20} cm^{-2}$ \citep{Murphy1996}. 

To obtain the X-ray spectrum for the particular time period individual observations were combined using \texttt{addspec} and the backgrounds were added together in \texttt{mathpha}. Further, the combined source and background spectra were added in \texttt{grppha} to tie together and the final SED was created in \texttt{Xspec}. The best fit results from 0.3-8 keV give a spectrum index $\Gamma_{X-ray}$ = 0.35$\pm$0.08. The X-ray flux found to be F$_{X-ray}$ = (1.07$\pm$0.06) $\times$10$^{-12}$ erg cm$^{-2}$ s$^{-1}$ with reduced chi-square, $\chi^2$/dof = 0.97 for 31 degrees of freedom. The X-ray light curve is shown in Figure~\ref{fig:LC}.

{\bf NuSTAR:}

\begin{figure}
\centering
\includegraphics[scale=0.36]{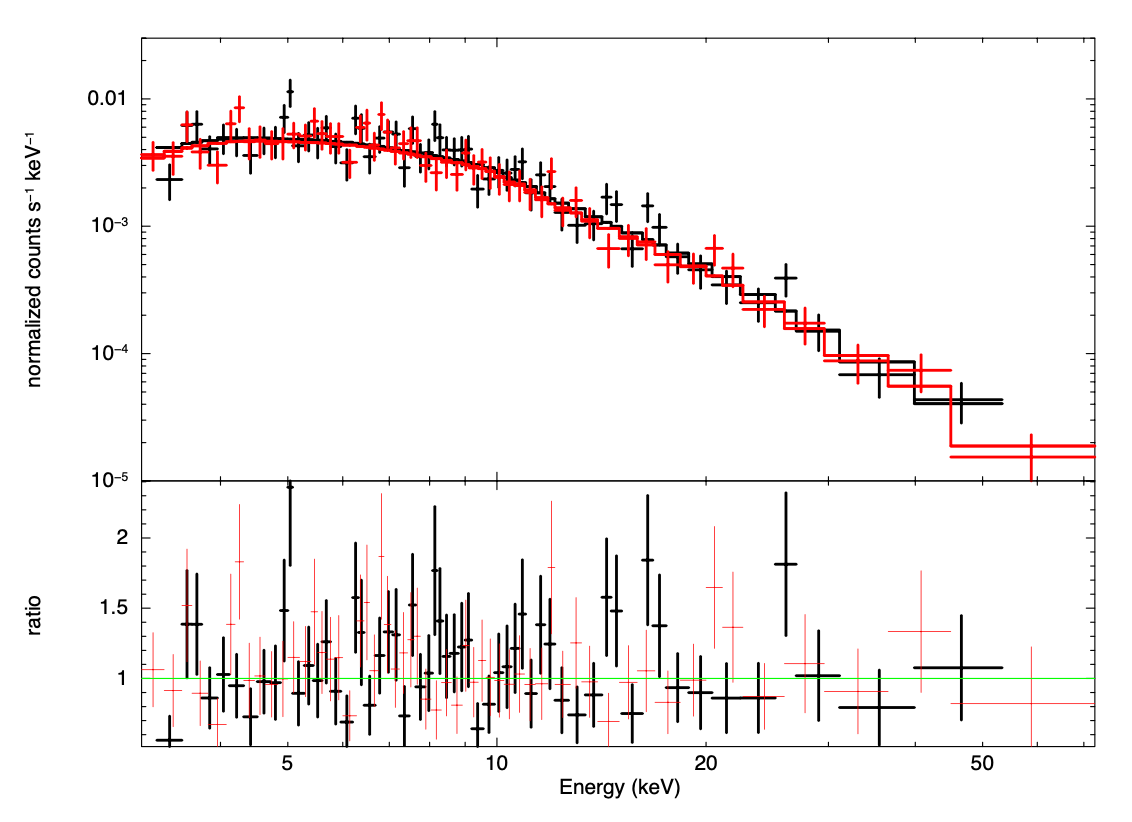}
\caption{NuSTAR spectrum taken on May 16, 2021. The best fit powerlaw spectrum for both the modules with black (FPMA) and red (FPMB) color}
\label{fig:nustar}
\end{figure}

NuSTAR is a hard X-ray telescope working in  3-79 keV energy band. It has two identical focal plane modules (FPMs) namely FPMA and FPMB. Blazar S5 1803+78 was observed May 16, 2021 (MJD 59350) for a total exposure of $\sim$20 ksec. The data was collected from HEASARC webpage and processed with NuSTAR data analysis software (NuSTARDAS) package using the latest calibration files. The collected raw data is cleaned with the \texttt{nupipeline} tools and clean event files are produced for both the modules (FPMA and FPMB). The source X-ray spectra for 3-79 keV were extracted using the circular region of size 25$''$ around the source.
A circular region of size 50$''$ was chosen away from the source location to avoid the source contamination for the background region. The source and background region files were used to create the nuSTAR products by running the tool \texttt{nuprodcuts}. The source and the background spectra were combined in \texttt{grppha} and binned to have at least 20 counts/bin. Eventually, the binned spectrum is fitted with an absorbed power law model in \texttt{Xspec}. The best fit results are following: The X-ray spectrum index for 3-79 keV is, $\Gamma_{X-ray}$ = 1.43$\pm$0.07 and the X-ray flux found as, F$_{X-ray}$ = (1.75$\pm$0.07) $\times$10$^{-11}$ erg cm$^{-2}$ s$^{-1}$ with reduced chi-square, $\chi^2$/dof = 0.93 for 111 degrees of freedom. The best fit power law spectrum for both the modules are shown in Figure \ref{fig:nustar} with black (FPMA) and red (FPMB) color. The spectrum is finally used in the broadband SED modeling in a later section.

\subsection{UV-optical observations of \textit{Swift}-UVOT}

The UV-optical observations from \textit{Swift}-UVOT \citep{swift2005} were used to get the multi-waveband light curves and the spectrum. 
The observations between MJD 58696-58957 were used in this work. The \textit{Swift}-UVOT obtains the data in six filters. Three optical filters in V, B, and U bands, and three UV filters, W1, M2, and W2. 
Only the U band among the optical filters was available for this source. The \texttt{uvotsource} tool was used to obtain the magnitude. This was corrected for reddening and galactic extinction using $E(B-V)=0.0448$ \citep{Schlafly_2011}. The corrected magnitudes were converted to flux using the \textit{Swift}-UVOT zero point magnitudes \citep{Giommi2006}. The UVOT light curve is shown in Figure~\ref{fig:LC}. The \texttt{uvotimsum} tool was used to sum all images from the individual filters and construct the UVOT SED.

\subsection{Optical observations }

The observations of the blazar were performed in the Johnson BVRI bands using the 0.6m RC robotic (T60) and the 1.0m RC (T100) telescopes at TUBITAK National Observatory in Turkey. The standard data reduction of all CCD frames, i.e. the bias subtraction, twilight flat-fielding, and cosmic-ray removal, was performed \citep{2019MNRAS.488.4093A}. A more detailed information on optical observations and their analysis will be published in the other upcoming paper (Agarwal et al.\ 2022, Under Review).

\subsection{Radio observations of OVRO}

Radio data for the source was obtained from the 40m telescope at the Owens Valley Radio Observatory (OVRO) \citep{Richard2011}. The observations were carried out at 15 GHz. The data was used to calculate the correlation with the Fermi-LAT data. The radio light curve is shown in Figure~\ref{fig:LC}.

\section{Multi-waveband light curve}
\label{sec3}
\begin{figure*}
    \centering
	\includegraphics[width=0.9\textwidth]{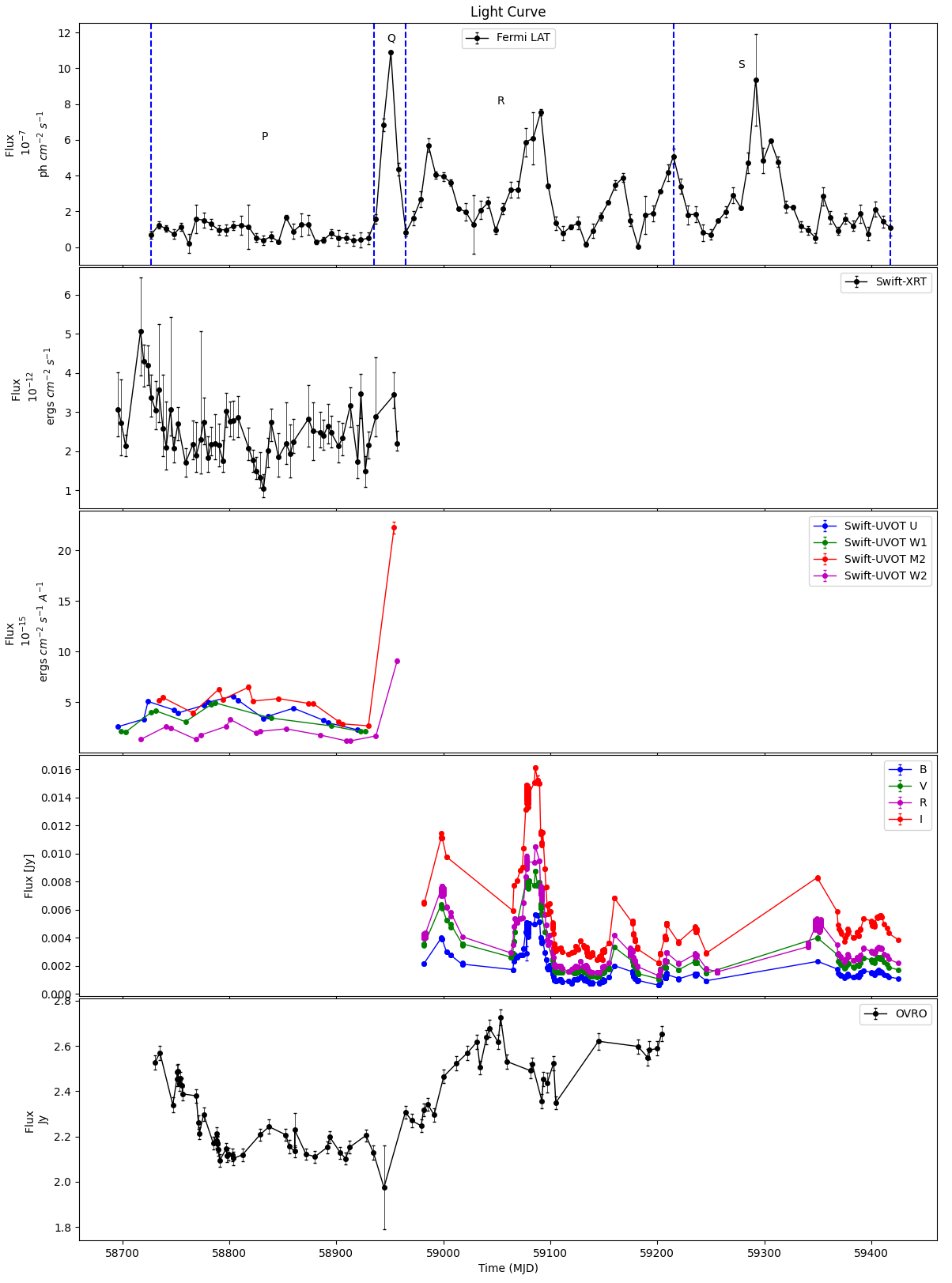}
    \caption{Multi-wavelength light curve for the source S5 1803+78}
    \label{fig:LC}
\end{figure*}

The multi-waveband light curve was obtained by combining data from the $\gamma$-ray, X-ray, UV-Optical, and radio wavelengths. We have divided the total time period into four distinct regions, depending upon their flux behavior, for spectral analysis which are called P, Q, R and S. P is defined as a pre-flaring region from MJD 58727 to MJD 58935 with an average $\gamma$-ray flux $0.88 \times 10^{-7} $ ph cm$^{-2}$ s$^{-1}$ from \textit{Fermi}-LAT , Q is a big flare observed from MJD 58935 to MJD 58965 with average $\gamma$-ray flux $4.90 \times 10^{-7} $ ph cm$^{-2}$ s$^{-1}$, R is the flaring region from MJD 58965 to MJD 59215 and the extended flaring region S from MJD 59215 to MJD 59419 with average $\gamma$-ray fluxes $2.70 \times 10^{-7} $ and $2.48 \times 10^{-7} $ ph cm$^{-2}$ s$^{-1}$, respectively. Swift-XRT and Swift-UVOT data were only available for the P and Q regions. The X-ray light curve does not show a simultaneous big flare in the Q region but in UV bands a signature of flare is seen simultaneously in the bands that were observed which are M2 and W2. The optical data in BVRI band from TUBITAK observatory show a simultaneous flare with the $\gamma$-ray data in the R region. The radio data at 15 GHz cover the P, Q and R regions, so include some flaring events.
The multi-wavelength light curve is shown in Figure~\ref{fig:LC}. The temporal and spectral study of various identified regions is presented in further sections.

\subsection{Variability time scale computation}
Characteristic variability timescale is an important parameter to understand the nature of flares and their production cite along the jet axis.
Doubling/Halving timescales are calculated for all time bins from MJD 58727 to 59420 for the $\gamma$-ray light curve. The formula used is as follows
\begin{gather}F(t_2)=F(t_1)\times 2^{(t_2-t_1)/T_d}\end{gather}

Here $F(t_1)$ and $F(t_2)$ are the fluxes measured at time $t_1$ and $t_2$, respectively. $T_d$ is the flux doubling/halving time scale. The fastest doubling/halving time ($T_{f}$) was found to be 1.37 days. The value for $t_{var}$ can be given by $t_{var} = ln(2)\times T_{f} $ which is 0.95 days. The day scale variability seen in this blazar suggests that the flares are produced close to the black hole within the broad-line region (BLR) under the shock in the jet model. The variability time can also be used to constrain the size of the emission region which we will discuss in Section 6.2.

To understand the structure of flares and their associated rise and decay timescale, we have fitted the total 1 day binned light curves with the sum of exponential.
The functional form of the sum of exponential is given by,

\begin{gather}f(t)=a_0+\sum_{i=1}^{N} 2a_i \Big[exp\Big(\frac{T_i-t}{T_{Ri}}\Big)+exp\Big(\frac{t-T_i}{T_{Di}}\Big)\Big]^{-1} \end{gather}

\begin{figure*}
    \centering
	\includegraphics[width=0.9\textwidth]{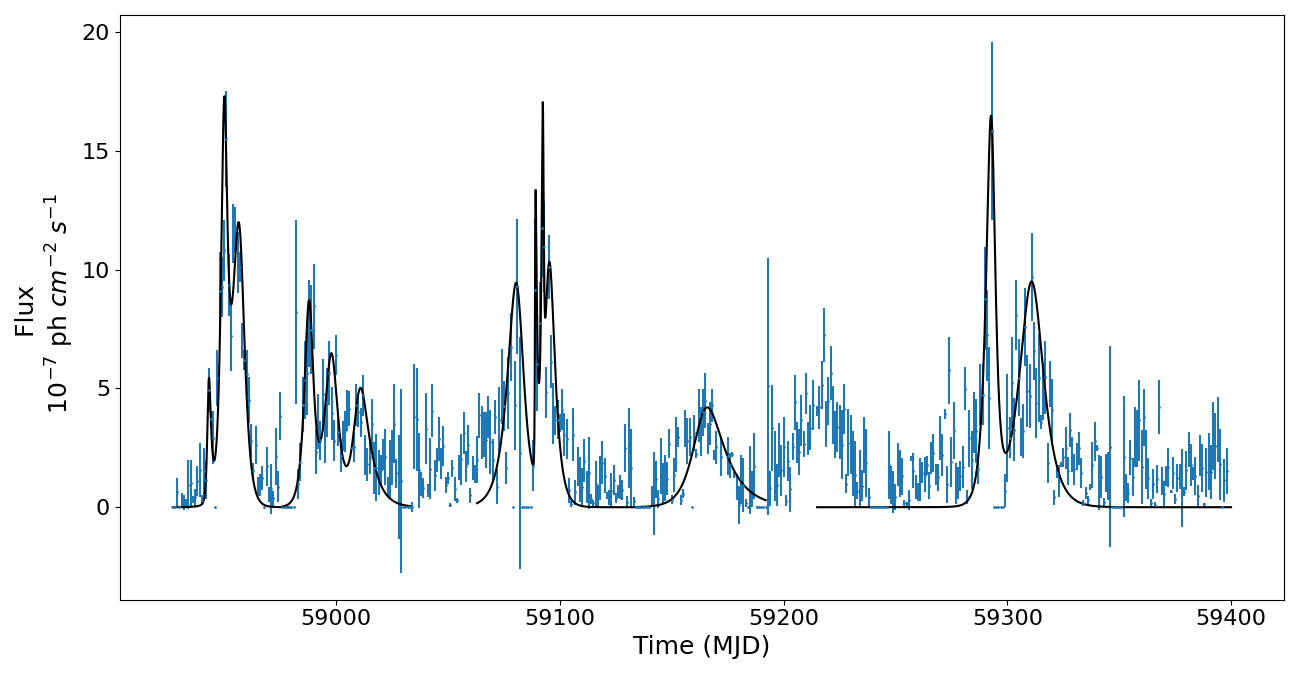}
    \caption{1-day binned light curve during the flaring episodes with the sum-of-exponentials fit  to obtain the rise and decay times.}
    \label{fig:fullfit}
\end{figure*}

Here $a_0$ is the baseline flux and the $a_i$s are the scaling factors which correspond to the peak flux. $T_{Ri}$ and $T_{Di}$ are the rise and decay times for the $i$th peak in the flare. $T_i$ gives the peak position for the $i$th peak. The fitting was done on the 1 day binned light curve which is shown in Figure~\ref{fig:fullfit}.

\begin{figure}
    \centering
	\includegraphics[width=\columnwidth]{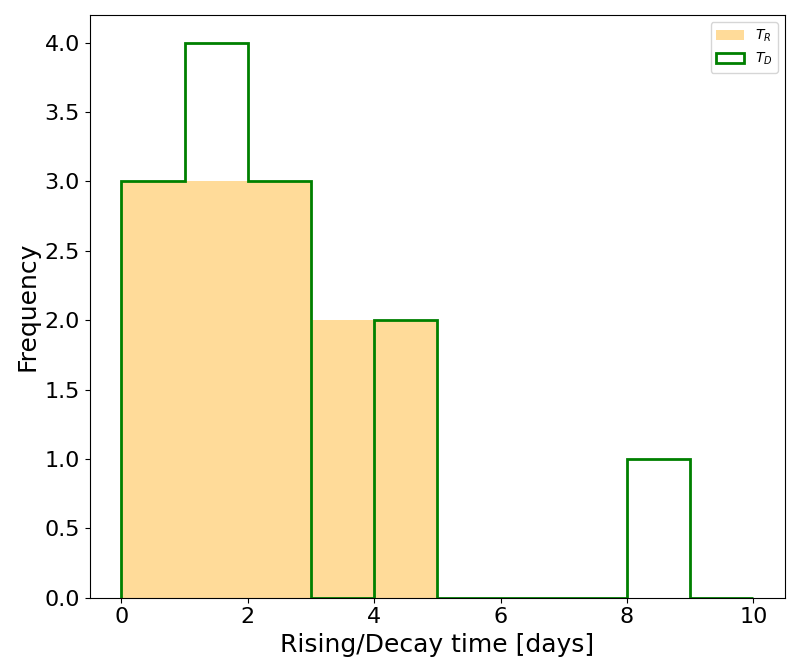}
    \caption{The distribution of the rise and decay times found from the sum-of-exponentials fit during the flaring states.}
    \label{fig:trtd}
\end{figure}

We have scanned the entire light curve including all the three flaring regions (Q, R, and S) with the above expression, and the fastest rise/decay time observed was 3.6 hours. The distribution of the rise and decay times, $T_R$ and $T_D$, can be seen in Figure~\ref{fig:trtd}, which suggests that most of the peaks in the flaring state have rise/decay time of the order of 1-2 days. This agrees with the variability time estimated independently in the above part as 0.95 days.

\subsection{Cross-correlation between different wavebands}
The location of gamma-ray flare in the jet of a blazar is still under debate. Various studies suggest that the day scale variability can be produced at the base of the jet under shock in the jet model but there are studies that have located the gamma-ray production site away from the BLR region to explain the fast flux variability in the mini jet model. The cross-correlation study among various wavebands can provide an indirect link to where the gamma-ray is produced and in general where the broadband emission is produced. It will also help to understand the time delay between various emissions that is mostly caused by the internal absorption in BLR or it can suggest the involvement of multi-zone emission along the jet axis.

Cross-correlation analysis was done between $\gamma$-ray and the different energy bands using the Discrete Correlation Function (DCF) as the data points are discrete and sampling is not even. The DCF allows computation of a correlation coefficient without using interpolation for data sampled at variable rates. 

The unbinned DCF function can be calculated for two data sets with data point ‘i’ in set 1 and data point ‘j’ in set 2 as \citep{Edelson1988}:

\begin{gather}DCF(\tau)=\frac{UDCF_{ij}}{M}\end{gather}

\begin{gather}
    UDCF_{ij}=\frac{(a_i-\bar{a})(b_i-\bar{b})}{\sqrt{(\sigma^2_a-e^2_a)(\sigma^2_b-e^2_b)}}
    \end{gather}

Here $M$ is the total number of pairs for which $(\tau - \Delta \tau/2) \leq \Delta t_{ij} \leq (\tau +\Delta\tau/2)$. $\Delta t_{ij} = (t_j-t_i)$ is the lag for the pair $(a_i,b_i)$. $\bar{a}$ and $\bar{b}$ are the average values of $a_i$ and $b_i$. The error for the DCF is given by,

\begin{equation}
\sigma_{DCF(\tau)}=\frac{1}{M-1}\sum_{i,j}(UDCF_{ij}-DCF(\tau))^{1/2}
\end{equation}

A significant correlation with positive time lag suggests that the first time series is leading the second time series and vice versa. In our analysis, the first time series is the $\gamma$-ray light curve. 

\begin{figure*}
    \centering
	\includegraphics[width=0.8\textwidth]{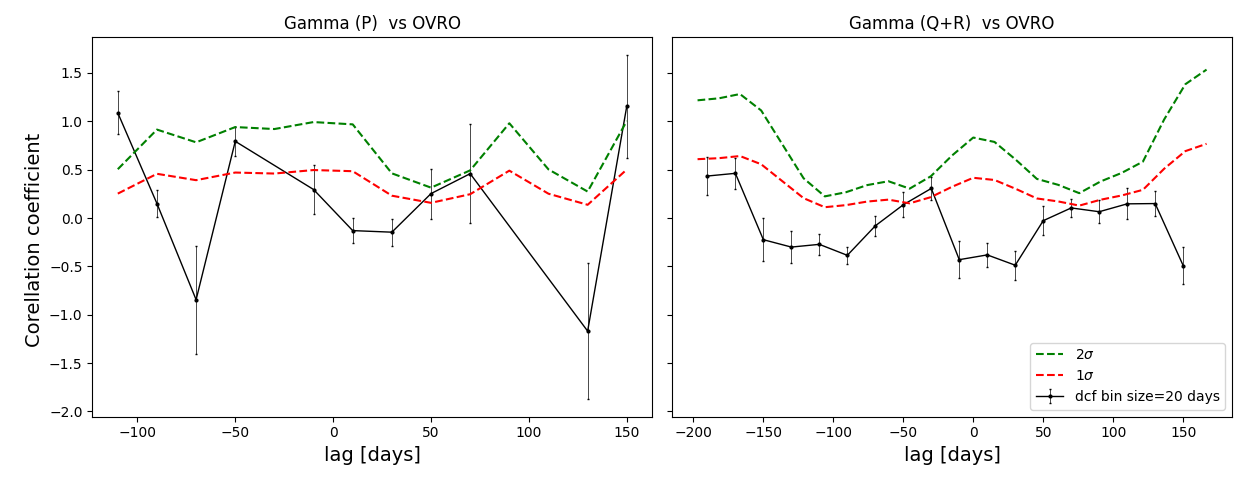}
    \caption{Correlation coefficient as a function of time lag for Gamma vs OVRO data}
    \label{fig:gmov}
\end{figure*}

\begin{figure}
    \centering
	\includegraphics[width=\columnwidth]{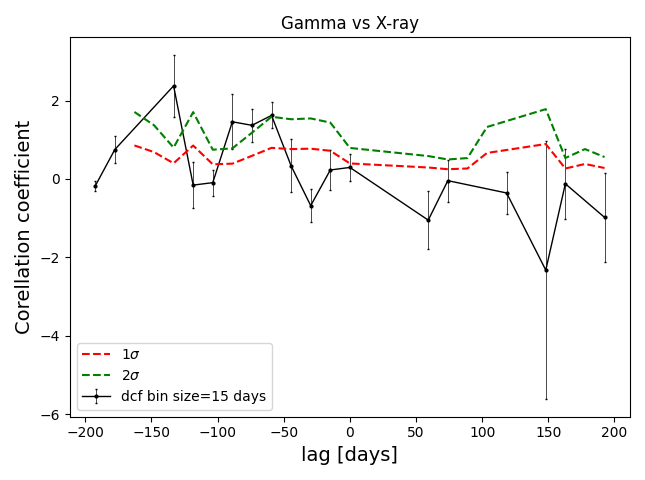}
    \caption{Correlation coefficient as a function of time lag for Gamma vs X-ray data}
    \label{fig:gmxr}
\end{figure}

The $\gamma$-ray light curve of the pre-flaring and the flaring region was cross-correlated with the OVRO light curve and the result is shown in Figure~\ref{fig:gmov}.
\citet{2002AJ....124...53N} have also shown that there is no significant correlation between the Medicina radio telescope at 8.4 GHz and optical emission in the
years 1996–2002. A similar result is also reported in \citet{Nesci2021} where no correlation is observed between OVRO 15-GHz and $\gamma$-ray emission.
They don't show any significant correlation. The DCF estimated for $\gamma$-ray light curve with the simultaneous X-ray light curve is shown in Figure~\ref{fig:gmxr}. No significant correlation was observed. Similar findings were also reported in \citet{Nesci2021} where the correlation between $\gamma$-ray and X-ray emission for this source is attempted. As expected, in BL Lac source as S5 1803+784, the X-ray and $\gamma$-ray emission are the part of SSC emission where the X-ray represents the lower part of the IC scattering and $\gamma$-ray at the higher part of the IC emission. In ideal situations, these emissions should show a clear correlation with their fluxes. However, the X-ray data are only available for the beginning of the light curve where the $\gamma$-ray is in a low state (region P) and steady throughout the period covered by the X-ray observation. Surprisingly x-rays show some variation in the flux during the same period but with a bigger error bar and hence no clear correlation between $\gamma$-ray and X-ray is concluded.

\begin{figure*}
    \centering
	\includegraphics[width=0.8\textwidth]{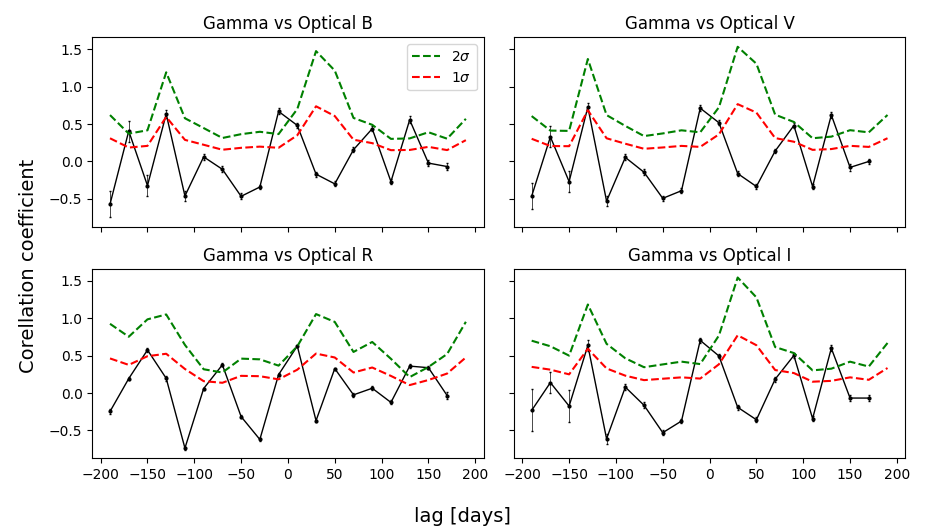}
    \caption{Correlation coefficient as a function of time lag for Gamma vs Optical BVRI data}
    \label{fig:gmopt}
\end{figure*}

The $\gamma$-ray flux is correlated (above $2\sigma$) with zero lag with the optical B, V, and I optical bands. We can see the DCF peak at zero time lag with significance greater than $2\sigma$ in Figure~\ref{fig:gmopt}. A positive correlation among optical and $\gamma$-ray has previously been observed by many authors \citep{Liodakis_2019} suggesting that both emissions are produced by the same population of electrons and are co-spatial. That means a single-zone emission region would be sufficient to explain the broadband SED which agrees with our result discussed in the SED modeling section.

To estimate the significance of the DCF, we simulated the 1000 artificial light curves by following the procedure outlined in \citet{Emma2013} which is documented in python code by \citet{Connolly2015} and available on \texttt{GitHub}\footnote{https://github.com/samconnolly/DELightcurveSimulation}. The simulated light curves were cross-correlated with the observed light curve. By collecting the time delay from each correlation a 1$\sigma$ and 2$\sigma$ significance is estimated and plotted in Figures~\ref{fig:gmov}, \ref{fig:gmxr}, \ref{fig:gmopt} in color red and green.  


\subsection{Colour magnitude}

\begin{figure}
   
    \centering
	\includegraphics[width=\columnwidth]{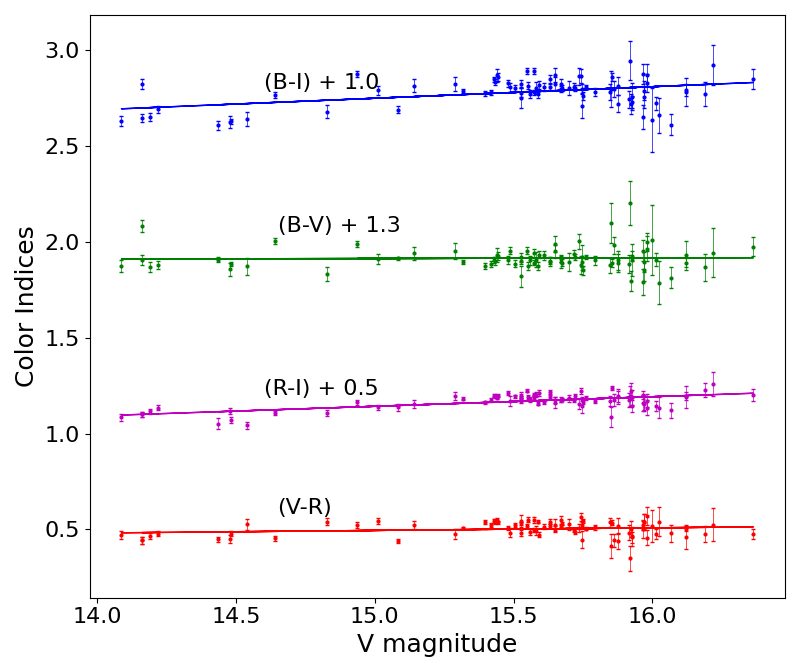}
    
    \caption{Colour-magnitude plot for S5 1803+78. The various colour indices are plotted against V band magnitude.}
     \label{mag}
\end{figure}

Analyzing the color-magnitude relationship can help us in understanding the origin of the blazar and probing the different variability scenarios. To obtain the relation between the color indices (CI) with respect to the V magnitude we fitted the plot with straight lines (CI $= mV +c$). The Spearman correlation coefficient $r$ was calculated along with the null hypothesis probability $p$. The fit values and the corresponding coefficients are presented in Table~\ref{tab:col mag}. A positive slope signifies a positive correlation and this is significant if $p \leq 0.05$ and $r \geq 0.6$. Such a relation tells us that the source displays a bluer when brighter (BWB) trend whereas a significant negative slope would mean a redder when brighter (RWB) trend \citep{hess2014}. The color-magnitude plot is shown in Figure~\ref{mag}. The offset values of 1.0, 1.3, and 0.5 are used with (B-I), (B-V), and (R-I) bands in Figure~\ref{mag} respectively for clarity. We have used inter-day observations. \citet{aditi2015} found a dominant BWB trend for their BL Lacertae source but we did not obtain a significant positive correlation between these indices and the V magnitude. The calculated fit values do not show a significant Spearman correlation coefficient for any of the indices, so no conclusive evidence can be drawn about the correlation from the available data.

\begin{table}
 \caption{Colour magnitude fitting and correlations coefficient. Here $r$ = Spearman coefficient and $p$ = null hypothesis probability}
 \label{tab:col mag}
 \begin{tabular}{lcccc}
  \hline
  Colour Indices & slope & intercept & $r$ & $p$\\
   
  \hline
  (B-I) & 0.085 $\pm$ 0.012 & 0.482$\pm$ 0.190 & 0.112 $\pm$ 0.057 & 0.320 \\
  (B-V) & -0.002 $\pm$ 0.008 & 0.643 $\pm$ 0.125 & 0.047 $\pm$ 0.080 & 0.674 \\
  (R-I) & 0.056 $\pm$ 0.006 & -0.184 $\pm$ 0.093 & 0.327 $\pm$ 0.064 & 0.003 \\
  (V-R) & 0.032 $\pm$ 0.007 & 0.022 $\pm$ 0.105 & 0.010 $\pm$ 0.075 & 0.931 \\
  \hline
 \end{tabular}
\end{table}

\section{gamma-RAY SPECTRAL DATA}
\label{sec4}
The following section outlines how the $\gamma$-ray spectral data from $Fermi$-LAT was created. The spectral energy distribution and the flux index relation is described below.

\subsection{Spectral model fits}

\begin{figure*}
    \centering
	\includegraphics[width=0.8\textwidth]{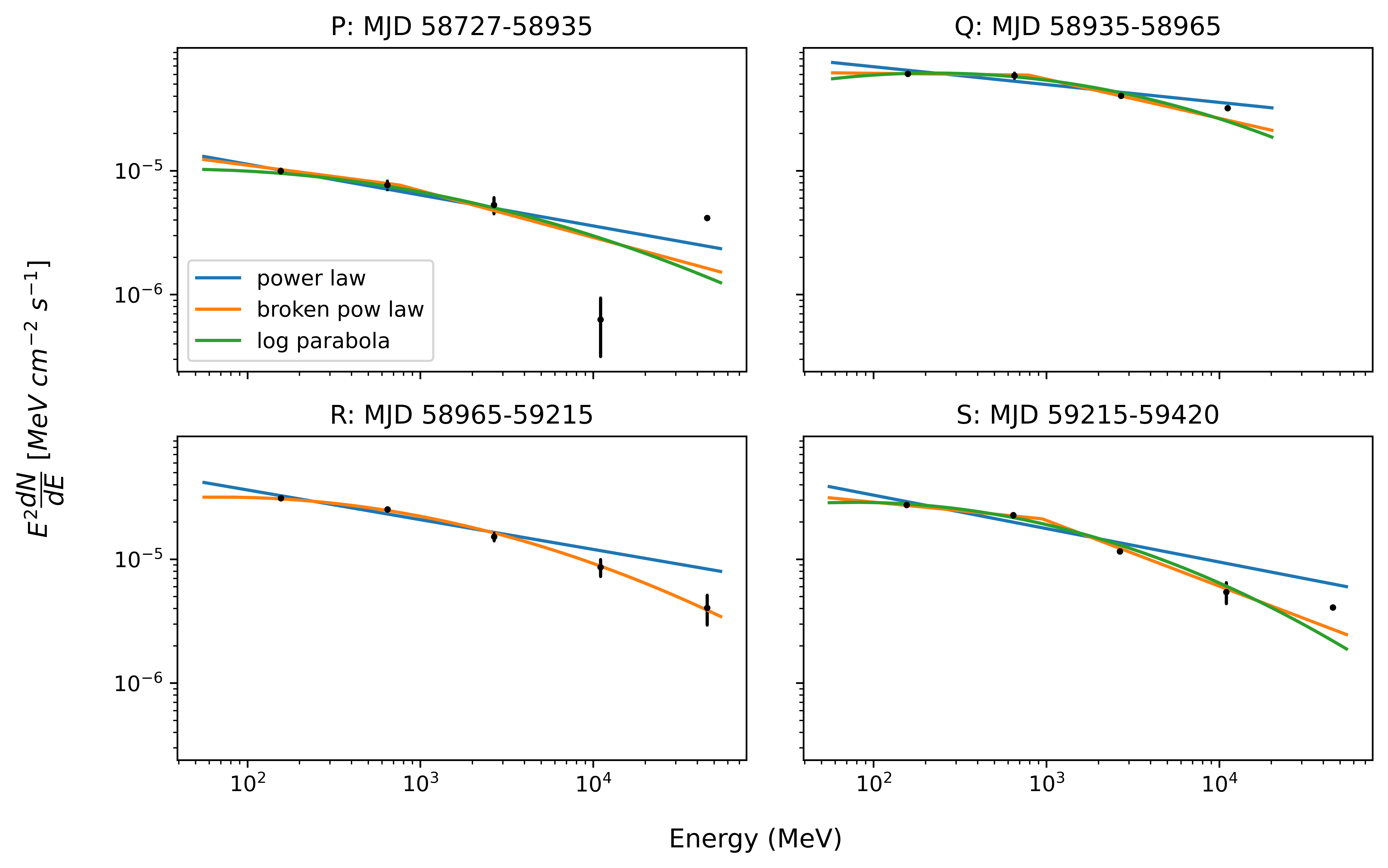}
    \caption{Comparison of spectral models fitted to $\gamma$-ray spectral energy distribution for the four regions P, Q, R and S.}
    \label{fig:sed2}
\end{figure*}

The $\gamma$-ray spectral analysis was done following the standard procedure in \texttt{Fermipy}. The SED was fitted with three spectral models, Power Law (PL), Broken Power Law (BPL) and Log Parabola (LP) that are described below,
\begin{gather} PL: \frac{dN(E)}{dE}=N_0 \times \bigg(\frac{E}{E_0}\bigg)^{-\alpha}\\
 BPL: \frac{dN(E)}{dE}= \begin{cases}
			N_0 \times (\frac{E}{E_0})^{-\alpha_1} , \; E<E_0\\
			N_0 \times (\frac{E}{E_0})^{-\alpha_2} , \; E \geq E_0
		 \end{cases}\\
 LP: \frac{dN(E)}{dE}=N_0 \times \bigg(\frac{E}{E_0}\bigg)^{-(\alpha+\beta log(E/E_0))}\end{gather}
In the above equations, $N_0$ is the prefactor, $E_0$ is the energy scale factor, $\alpha$, $\alpha_1$ and $\alpha_2$ are spectral indices and $\beta$ in log parabola is the curvature index.
The likelihood for the BPL model could not converge for the flaring region R and hence no successful $\gamma$-ray SED fitting is done.  The comparison of these models for all the four regions is shown in Figure~\ref{fig:sed2}. The fitted parameters are presented in Table~\ref{table:sedpt}. The likelihood analysis also produces the TS value corresponding to each fit which describes the goodness of the fit. Based on just the TS value it is very difficult to make any strong conclusion about the best fit model to the $\gamma$-ray SED. However, both the LP and BPL are the best candidate to explain the $\gamma$-ray SED. As noted in the PL case, during the region P the spectrum is 2.25 and became a bit harder with index 2.14 during region Q but again became softer during region R and S. Suggesting both harder when brighter trend in the beginning and softer-when-brighter for the brightest flare. As reported by \citet{Nesci2021}, the average $\gamma$-ray spectral index for low and flaring states are 2.22$\pm$0.01 and 2.21$\pm$0.02 which agrees with our spectral fitting also where the region P (low state) has an average spectral index 2.25 and the brightest flaring state (region R) has average spectral index 2.24 which are described by the same spectral index.

\begin{table*}
\caption{Best fit parameters for different spectral models for $\gamma$-ray data.}
\label{table:sedpt}
\centering

\subcaption{Power Law}
\begin{tabular}{p{0.05\linewidth}p{0.15\linewidth}p{0.1\linewidth}p{0.1\linewidth}p{0.1\linewidth}p{0.1\linewidth}p{0.1\linewidth}}
\hline
Region & Flux [ph cm$^{-2}$ s$^{-1}$] & $N_0$ & $E_0$ [MeV] & $\alpha$ & TS \\
\hline
P & $0.9 \times 10^{-7}$  & 1.7 &  645 & 2.25 & 916.24\\
Q & $6.0 \times 10^{-7}$ &  12.6 & 646 & 2.14 & 3326.13\\
R & $2.9 \times 10^{-7}$ & 5.5 & 646 & 2.24 & 7027.01\\
S & $2.6 \times 10^{-7}$ &  4.8 & 645 & 2.27 & 5320.26\\
\hline
\end{tabular}

\subcaption{Broken Power Law}
\begin{tabular}{p{0.05\linewidth}p{0.15\linewidth}p{0.1\linewidth}p{0.1\linewidth}p{0.1\linewidth}p{0.1\linewidth}p{0.1\linewidth}}
\hline
Region & Flux [ph cm$^{-2}$ s$^{-1}$] & $N_0$ & $E_0$ & $\alpha_1$ & $\alpha_2$ & TS \\
\hline
P & $0.86 \times 10^{-7}$ & 0.013 & 761 & -2.13 & -2.38 & 916.76\\
Q & $5.9 \times 10^{-7}$ &  0.094 & 792 & -2.02 & -2.32 &  3321.63\\
S & $2.5 \times 10^{-7}$ &  0.024 & 948 & -2.14 & -2.53 & 5327.55\\
\hline
\end{tabular}

\subcaption{Log Parabola}
\begin{tabular}{p{0.05\linewidth}p{0.15\linewidth}p{0.1\linewidth}p{0.1\linewidth}p{0.1\linewidth}p{0.1\linewidth}p{0.1\linewidth}}
\hline
Region & Flux [ph cm$^{-2}$ s$^{-1}$] & $N_0$ & $E_0$ & $\alpha$ & $\beta$ & TS \\
\hline
P & $0.87 \times 10^{-7}$ & 1.80 & 645 & 2.23 & $4.00 \times 10^{-2}$ & 916.46\\
Q & $5.8 \times 10^{-7}$ & 1.38 & 645 & 2.12 & $5.85 \times 10^{-2}$ &  3321.65\\
R & $2.8 \times 10^{-7}$ & 5.94 & 645 & 2.22 & $4.98 \times 10^{-2}$ & 6998.34\\
S & $2.5 \times 10^{-7}$ & 5.20 & 645 & 2.27 & $6.34 \times 10^{-2}$ & 5358.11\\
\hline
\end{tabular}



\end{table*}

\subsection{Flux Index correlation}

\begin{figure*}
    \centering
	\includegraphics[scale=0.35]{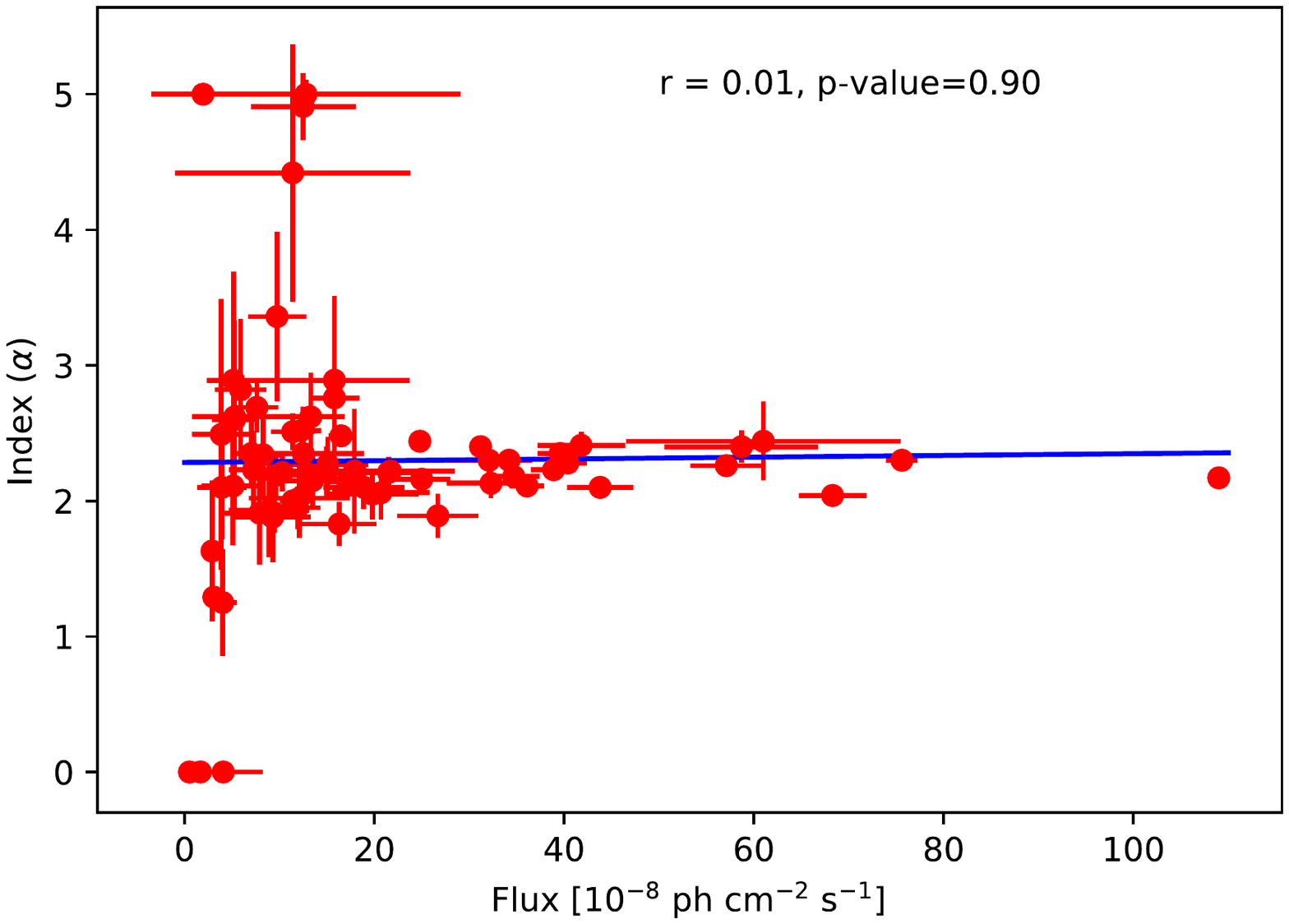}
	\includegraphics[scale=0.35]{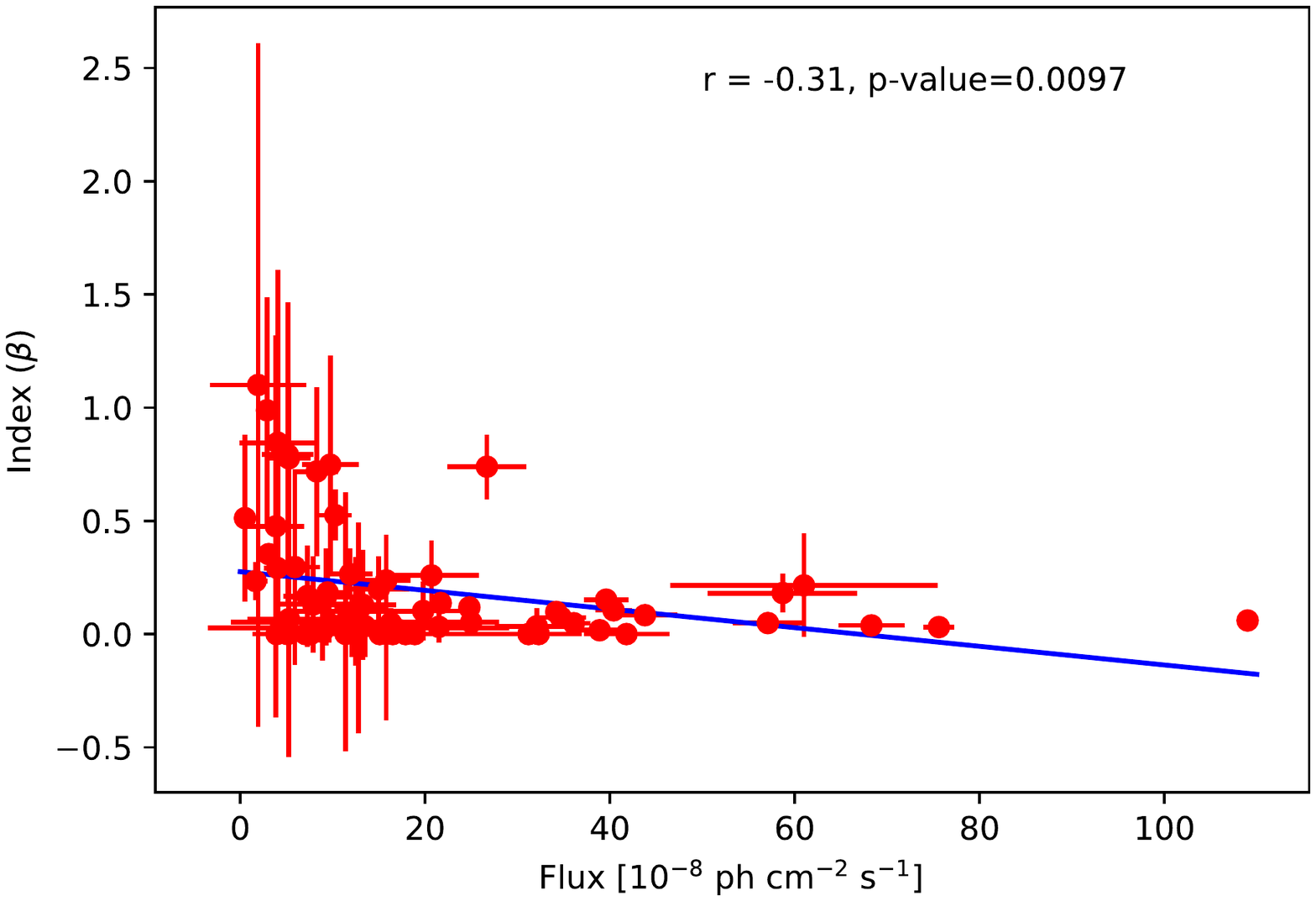}
    \caption{Flux vs index plots for \textit{Fermi}-LAT data. The Pearson coefficient correlation and the p value have been mentioned in the plot for both indices.}
    \label{fig:fic}
\end{figure*}

The flux-index correlations were computed for the $\gamma$-ray lightcurve and the plot is shown in Figure~\ref{fig:fic}. It shows that the flux does not have a significant correlation with the spectral index, $\alpha$. The Pearson correlation coefficient is 0.01 with a p-value of 0.90. A slight anti-correlation is observed between flux and the curvature index, $\beta$ with Pearson correlation coefficient -0.31 and a p-value of 0.0097 which suggests a brighter-when-softer trend. However, no such trend is observed with $\alpha$. No clear trend in flux-index relation is also seen by \citet{Nesci2021}, however, when they consider only higher significance flux points (TS$\geq$25) a positive correlation is observed. However, in some of the previous study on FSRQ and BL Lac type sources, different trend such as harder-when-brighter and softer-when-brighter trend is seen (\citealt{Prince2019, Prince2020, Prince2021}).


\section{multi-wavelength SED Modeling}
The broadband SED modeling in blazar is important to understand the simultaneous or quasi-simultaneous multi-wavelength emission in the source along with the possible physical mechanism responsible for broadband flaring event. In BL Lac type sources such as S5 1803+78, a one-zone synchrotron-self Compton (SSC) model is expected to explain the broadband emission. In this scenario, the charged particle motion in the magnetic field produces the synchrotron photons, and these synchrotron photons can get up-scatter by the same electrons to very high energy through the inverse-Compton (IC) process and produces the gamma-ray flare followed by the lower frequency emission such as X-ray and optical-UV, such model is also known as leptonic model. 

In this paper, we have carried out broadband SED modeling for the pre-flaring period P and for the three flaring episodes denoted by Q, R \& S identified in the Figure~\ref{fig:LC}. The modeling is done using the publicly available code JETSET\footnote{https://jetset.readthedocs.io/en/latest/} \citep{Massaro2006, Tramacere2009, Tramacere2011, Tramacere2020}. The model assumes that the emission region is a spherical blob of radius R, filled with relativistic electrons, magnetized with a field strength B and it moves along the jet with the bulk Lorentz factor of $\Gamma$. The angle of inclination with the observer is $\theta$. The electron population is described by a broken power-law \citep{Katarzy2001}.

\begin{equation}
N(\gamma)=\begin{cases}
			C_1 \times \gamma^{\alpha_1} \; , \; \gamma_{min} \leq \gamma \leq \gamma_{break}\\
			C_2 \times \gamma^{\alpha_2} \; , \; \gamma_{break} \leq \gamma \leq \gamma_{max} 
		 \end{cases}
\end{equation}

Here $\gamma = E/mc^2$ is the Lorentz factor and $\alpha_1$ and $\alpha_2$ are the spectral indices before and after the break energy. $C_2$ and $C_1$ are normalization constants that satisfy the condition $C_2=C_1(\gamma_{break})^{\alpha_2-\alpha_1}$.

In the SED, the lower energy component between the radio to soft-X-ray or the first hump is due to the synchrotron photons produced by the interaction of relativistic electrons with the magnetic field. The second hump at higher energies between X-ray to $\gamma$-ray is produced via inverse Compton scattering. There are two ways this can happen either by SSC (Synchrotron-Self Compton) or by EC (external Compton). In the first case relativistic electrons up-scatter the same synchrotron photons which they have produced in the magnetic field. In the second case, the electrons can up-scatter external photons originated from 1) accretion disk, 2) from BLR (Broad Line Region) or 3) from the dusty torus. In BL Lac type sources, there is no evidence of the presence of BLR and hence most of the emission is thought to be produced by the SSC process.   

\begin{figure*}
    \centering
	\includegraphics[width=0.8\textwidth]{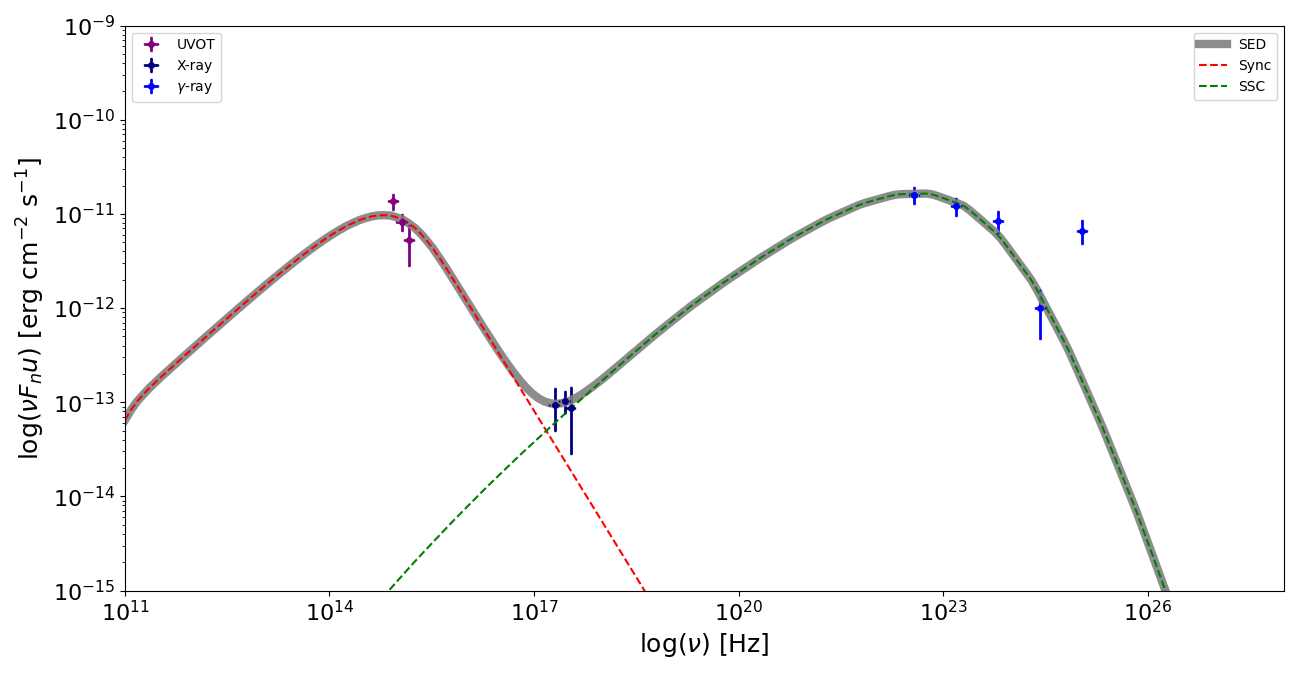}
    \caption{{\bf Spectral energy distribution of the region P. The observed SED is fitted with a simple one-zone leptonic model using the synchrtoron and SSC mechanism.}}
    \label{fig:sed1.1}
\end{figure*}

\begin{figure*}
    \centering
	\includegraphics[width=0.8\textwidth]{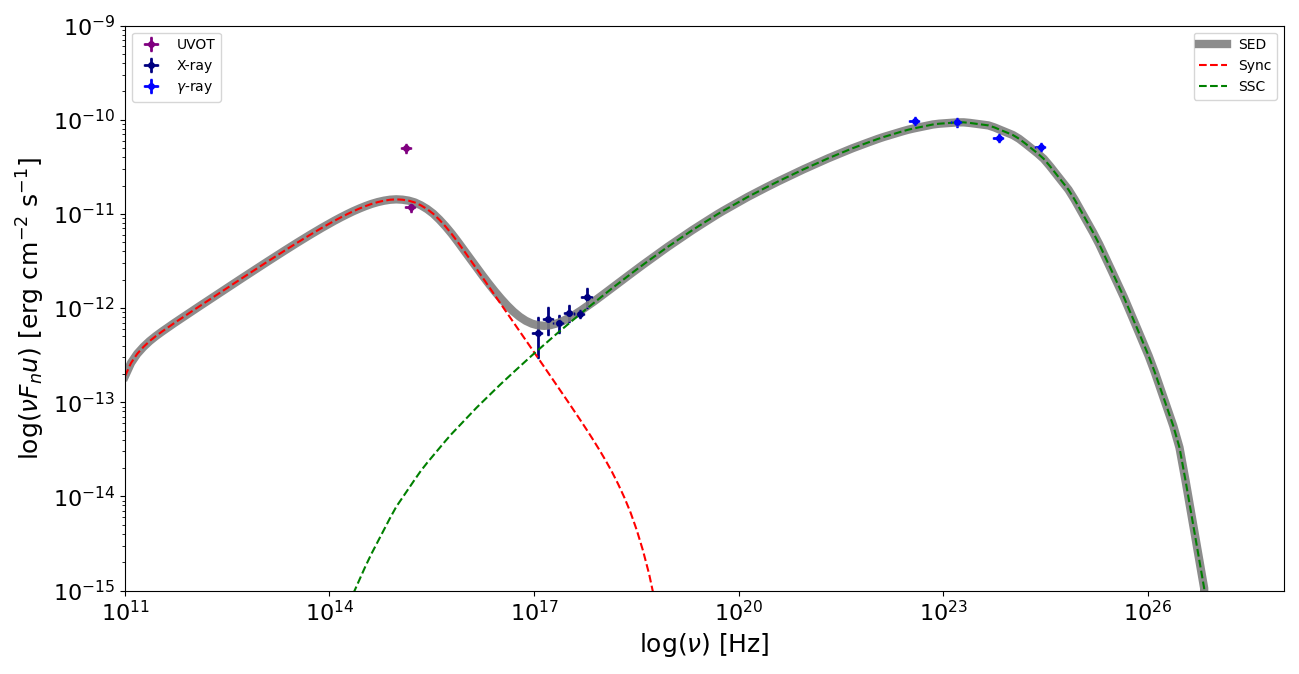}
    \caption{{\bf same as Figure 11.}}
    \label{fig:sed1.2}
\end{figure*}

\begin{figure*}
    \centering
	\includegraphics[width=0.8\textwidth]{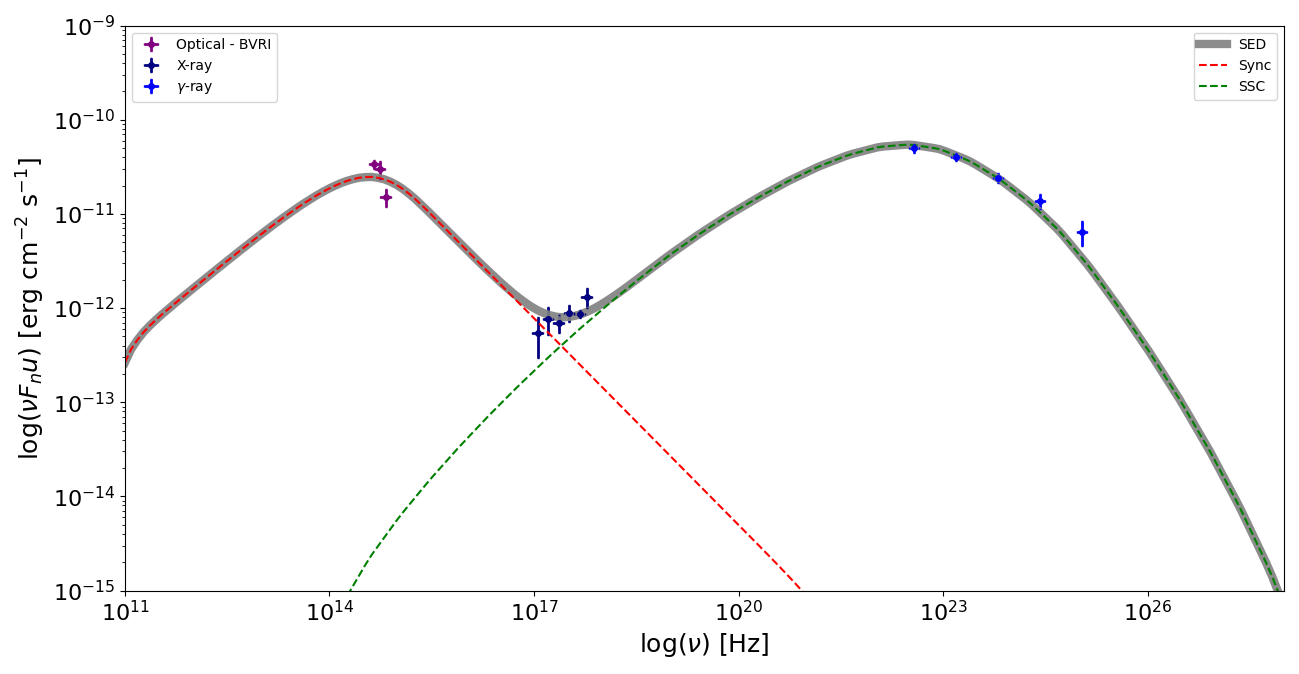}
    \caption{Spectral energy distribution of the observed data in region R with the fitted model. The X-ray data used here is from the Q region.}
    \label{fig:sed1.3}
\end{figure*}

\begin{figure*}
    \centering
	\includegraphics[width=0.8\textwidth]{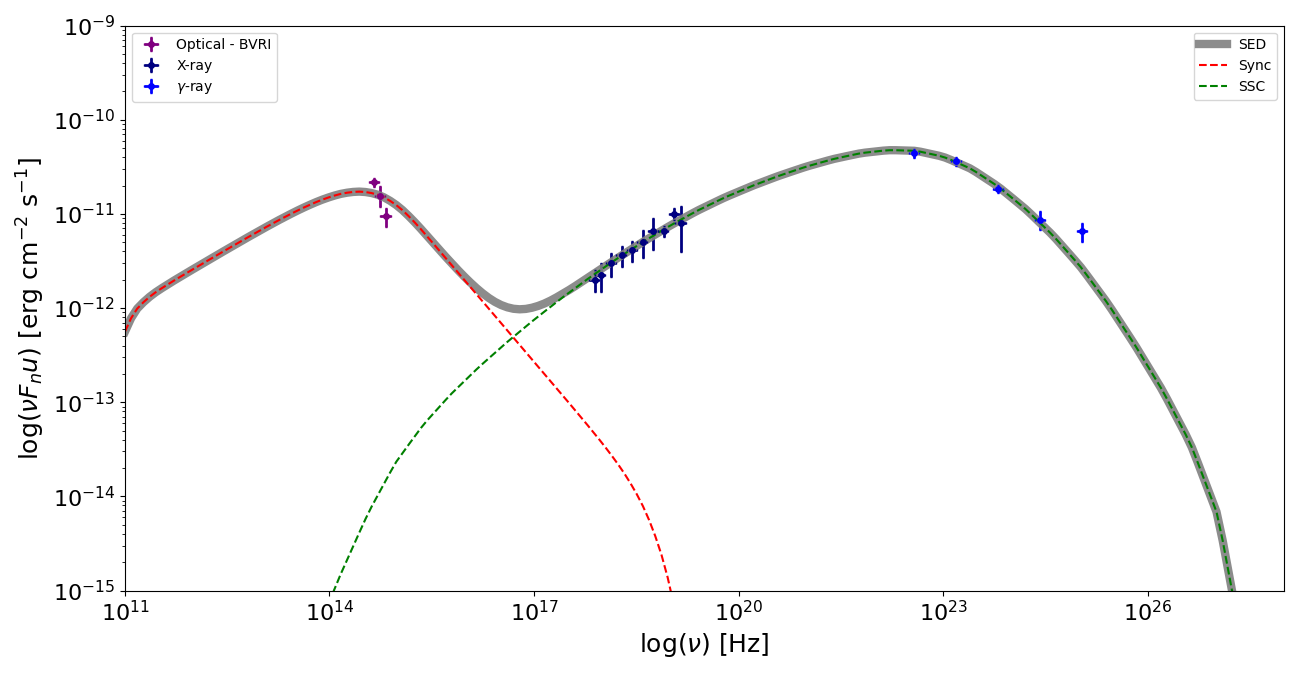}
    \caption{Spectral energy distribution of the observed data in region S with the fitted model.}
    \label{fig:sed1.4}
\end{figure*}

Blazar S5 1803+78 is identified as a BL Lac type source and hence
the modeling for S5 1803+78 is based on the SSC model. Figure~\ref{fig:sed1.1}, Figure~\ref{fig:sed1.2}, Figure~\ref{fig:sed1.3}, Figure~\ref{fig:sed1.4} show the modeled SED for the regions P, Q, R and S, respectively. The model parameters are given in Table~\ref{table:sed}. There was no simultaneous X-ray data available for the R region and we have used Swift-XRT data from the Q region to model the multi-wavelength SED for this region.

We calculated the jet power along with the power of individual components, 
i.e., leptons, protons and magnetic field. The total power of the jet is obtained using
\begin{equation}
    P_{jet}=\pi R^2 \Gamma^2 c (U_e^{'} + U_p^{'} + U_B^{'})
\end{equation}
Here $\Gamma$ is the bulk Lorentz factor. $U_e^{'}$, $U_p^{'}$, $U_B^{'}$ are the energy densities of electrons-positrons, cold protons and the magnetic field respectively in the co-moving jet's frame. The primed quantities are in the co-moving jet frame and the unprimed quantities are in the observer's frame. The energy densities for different components for all the four regions was returned by our model. We calculated $P_e$, $P_p$ and $P_B$ which are the power carried by the leptons, cold protons and the magnetic field respectively. The total power along with the power of the individual components has been mentioned in Table~\ref{tab:jetpow}. The jet is dominated by the leptons power and its value increases as the source travel from low flux state (P-region) to high flux states.

We have noticed that as the source moves from pre-flaring state (P: low flux state) to flaring state (high flux state)
 the Doppler factor changes from 14.76 in P state to 22.09 in the brightest state. The minimum and maximum energy of electrons are also increased from low state to high state suggesting high energy electrons are involved in producing the brightest $\gamma$-ray flares. The best fit magnetic field through the various states are almost the same and the value is of the order of 0.1 Gauss. Depending upon the shape of the broadband SED the spectral index before and after the break have been optimized.
 Overall the break energy of the electron distribution does not show any clear trend throughout the various regions/states.
 With our modeling results, we concluded that the various flaring states can be caused by the increase in the Doppler factor. Sudden increment in the Doppler factor can be linked to the sudden change in accretion disk assuming there is a disk-jet connection. Though this explanation is not unique there could be more complex situations that need to be explored in the future.

\begin{table*}
\caption{Best fit parameters for the SED modelling of the regions P, Q, R and S}
\label{table:sed}
\centering
\begin{tabular}{p{0.05\linewidth}p{0.15\linewidth}p{0.1\linewidth}p{0.1\linewidth}p{0.1\linewidth}p{0.1\linewidth}p{0.1\linewidth}}
\hline
Sr. No. & Model parameters & Unit & P & Q & R & S\\
\hline
1. & $R$ & $10^{16}$cm & 5.0 & 5.0 & 5.0 & 5.0\\
2. & $R_H$ & $10^{17}$cm & 1.0 & 1.0 & 1.0 & 1.0\\
3. & $B$ & G & 0.21 & 0.10 & 0.13 & 0.21\\
4. & $N$ & cm$^{-3}$& 153 & 550 & 153 & 710\\
5. & $\gamma_{min}$ & - & 9.83 & 24.14 & 15.59 & 4.11\\
6. & $\gamma_{max}$ & - & $5.5 \times 10^5$ & $5.0 \times 10^5$ & $8.2 \times 10^7$  & $2 \times 10^6$\\
7. & $\gamma_{break}$ & - & $9.9 \times 10^3$ & $1.8 \times 10^4$ & $7.2 \times 10^3$ & $3.9 \times 10^3$\\
8. & $\alpha_1$ & - & 1.74 & 2.01 & 1.80 & 1.66\\
9. & $\alpha_2$ & - & 5.43  & 5.10 & 4.46 & 3.88\\
10. & $z$& - & 0.68 & 0.68& 0.68 & 0.68\\
11. & $\delta$ & - & 14.76 & 16.41 & 22.09 & 16.16\\
\hline
\end{tabular}
 \vspace{1ex}

{\raggedright Note: [1] The size of the emission region; [2] The position of the region; [3] Magnetic field; [4] Particle density; [5-7] Minimum, maximum and the break Lorentz factor of the injected electron spectrum; [8] Low energy spectral index; [9] High energy spectral index; [10] Redshift; [11] The Doppler beaming factor. \par}
\end{table*}

\begin{table*}
 \caption{Total jet power with the breakdown of individual components}
 \label{tab:jetpow}
 \begin{tabular}{lllllll}
  \hline
  Power of the Jet &  & Unit & P & Q & R & S\\
   
  \hline
  Power in leptons & $P_e$  & $10^{42}$ ergs s$^{-1}$ & 754.28 & 3976.49 & 1001.41 & 1784.80 \\
  Power in cold protons & $P_p$ & $10^{42}$ ergs s$^{-1}$ & 1.11 & 0.05 & 0.29 & 0.03 \\
  Power in the Magnetic field & $P_B$ & $10^{42}$ ergs s$^{-1}$ & 55.13 & 5.28 & 4.13 & 0.40\\
  Total power & $P_{jet}$ & $10^{42}$ ergs s$^{-1}$ & 810.52 & 3981.82 & 1005.83 & 1785.23\\
  \hline
 \end{tabular}
\end{table*}

\section{Results and discussion}

\subsection{Minimum Doppler factor}
The minimum value of the Doppler factor can be calculated using the detection of the highest energy photons of the source. The optical depth, $\tau_{\gamma \gamma}$ of the highest energy photon to $\gamma \gamma$ interaction is assumed to be 1. The doppler factor then is given by the formula

\begin{equation}
\delta_{min} = \bigg[\frac{\sigma_t d_l^2 (1+z)^2 f_{\epsilon E_h}}{4t_{var} m_e c^4}\bigg]^{1/6}
\end{equation}

Here $\sigma_t$ is the Thomson cross-section for electrons, $d_l$ is the luminosity distance, $f_{\epsilon}$ is the X-ray flux and $E_h$ is the highest energy photon. The $d_l$ for this source is 4.16 Gpc and the observed variability timescale, $t_{var}$ was calculated to be 0.95 days. The highest-energy photon in the Q region was 11.17 GeV. Using the X-ray flux at this point the value of $\delta_{min}$ was found to be 6.8. Based on the radio observation from high-resolution VLBA images from the MOJAVE program and millimeter-wavelength flux density monitoring data from Metsähovi Radio Observatory \citet{Savolainen2010} have estimated the Doppler factor as 12.1 similar to \citet{Hovatta2009} but higher than the minimum Doppler factor estimated in our case. However, in our SED modeling, we estimated the Doppler factor between the range of 14.76--22.09 depending upon their flux state.
A low value of $gamma$-ray Doppler factor, 5.8,  was also estimated by \citet{Fan_2013}. Based on the multi-wavelength monitoring of Fermi blazar \citet{Liodakis2017} have derived the Doppler factor for this source and the reported value is 21.2 and which agrees with the Doppler factor estimated from our SED modeling with a minor difference.

\subsection{The emission region}
The information about the size and the location of the emission region along the jet axis is important to understand how the broadband emission is produced and what are the physical mechanisms involved. The fast-flux variability produced at the base of the jet (within the BLR) is most probably caused by the shock in the jet model. However, the minute scale variability produced at higher distances can be produced by the mini-jet model under the influence of magnetic reconnection as shown in \citet{Shukla2020}.

The variability time scale estimated from the $\gamma$-ray light curve is used to estimate the size of the emission region. The radius R can be estimated by the equation
\begin{equation} R = c \delta_{min} t_{var} / (1+z)
\end{equation}
Using the calculated variability time scale and the minimum Doppler factor, the size of the region was estimated to be $1.0 \times 10^{16}$ cm. For the SED modeling, we have used the value of $5 \times 10^{16}$ cm. The location of the emission region can be estimated by the expression, d $\sim$ 2c$\Gamma^2$ t$_{var}$/(1+z). Using the Lorentz factor, $\Gamma$ = 9.4 (\citealt{Savolainen2010}) and variability time 0.95 days (this paper) and z= 0.684, the  location is estimated to be, d $\sim$ 2.6$\times$10$^{17}$ cm, which is not very far from the central supermassive black hole. To optimize the broadband SED modeling, we have fixed the location of the emission region to 1$\times$10$^{17}$ cm along the jet axis.

\subsection{Broadband emission during flaring states}

A major flare occurred in S5 1803+78 during April 2020 followed by more flaring episodes. The highest flux observed during this period was $1.1 \times 10^{-6}$ ph cm$^{-2}$ s$^{-1}$ while in the pre-flaring region the flux was below $0.2 \times 10^{-6}$ ph cm$^{-2}$ s$^{-1}$. Figure~\ref{fig:sed1.1}, Figure~\ref{fig:sed1.2}, Figure~\ref{fig:sed1.3}, and Figure~\ref{fig:sed1.4} show the modelled SEDs for the pre-flaring and all the flaring periods respectively. Table~\ref{table:sed} shows the best fit model parameters. The model parameters for the regions are of similar order with minor differences. The low energy spectral index $\alpha_1$ ranges from 1.66 to 2.01 while the high energy index $\alpha_2$ goes from 3.88 to 5.43. We estimated the minimum Doppler factor as roughly $\delta_{min} = 6.8$ and we kept the parameter-free for a better fit. The magnetic field $B$ takes a value between 0.10$\sim$0.21. Synchroton emission from the jet dominated the Optical-UV part of the emission and the X-ray emission was reproduced by the SSC model. 

\section{SUMMARY}

The source S5 1803+78 was found to be in a flaring period from March 27, 2020 (MJD 58935) till July 24, 2021 (MJD 59419). We divided our data into four distinct regions with respect to flaring episodes and the availability of simultaneous observations in other bands. We have analyzed the flaring episodes along with the pre-flaring region since September 1, 2019. The fastest variability timescale was found to be 0.95 days from analyzing the light curve. The $\gamma$-ray emission was correlated with the different optical bands with zero lag and no significant correlation between $\gamma$-ray, radio, and X-ray is found. The source exhibits a bluer when brighter (BWB) trend which we concluded from the significant positive correlation between the (R-I) index and the V band magnitude. The highest-energy photon found during the big flare denoted as Q was 11.17 GeV. We compared different models and used the log parabola model for the $\gamma$-ray spectral data. The flux doesn't show any significant correlation with the spectral index and there is only a minor anti-correlation with the curvature index. The minimum Doppler factor was calculated to be 6.8 and the estimated emission region was found to be $1.0 \times 10^{16}$ cm. The multi-wavelength SED has been modeled for all the regions and the model parameters given in Table~\ref{table:sed} show a similar nature for the different episodes. Our SED modeling suggests that an increase in the Doppler factor is required to explain the flaring state from the low flux state which can be speculated as to the sudden change in the accretion rate of the accretion disk powering the jet.

\section*{Acknowledgements}
Authors thank the anonymous referee for their constructive and insightful comments and suggestions.
DB acknowledges Science and Engineering Research Board - Department of Science and Technology for Ramanujan Fellowship - SB/S2/ RJN-038/2017. RP acknowledges the support by the Polish Funding Agency National Science Centre, project 2017/26/A/ST9/00756 (MAESTRO 9), and MNiSW grant DIR/WK/2018/12. AA and AO were supported by The Scientific and Technological Research Council of Turkey (TUBITAK), through project number 121F427. We thank TUBITAK National Observatory for partial support in using T60 and T100 telescopes with project numbers 19BT60-1505 and 19AT100-1486, respectively. We also thank  Atatürk University Astrophysics Research and Application Center (ATASAM) for partial support in using the ATA50 telescope which was provided by Scientific Research Projects Coordination Units in Atatürk University (P.No. BAP-2010/40).

\section*{Data Availability}
For this work we have used data from the Fermi-LAT, Swift-XRT, Swift-UVOT, and NuSTAR which are available in the public domain. We have also used optical data collected by TUBITAK telescope and radio data seen by OVRO. These optical and radio data were given to us on request. Details are given in Section 2.




\bibliographystyle{mnras}
\bibliography{example} 








\bsp	
\label{lastpage}
\end{document}